# Research on signalized intersection mixed traffic flow platoon control method considering Backward-looking effect

Binghao Feng[a], Hui Guo[b*], Minghui Ma[c*], Yuepeng Wu[d], Shidong Liang[e], Yansong Wang[f]

[a] M.Eng, School of Mechanical and Automotive Engineering, Shanghai University of Engineering Science, Shanghai 201620, China (E-mail: m310122414@sues.edu.cn)

[b] Professor, School of Mechanical and Automotive Engineering, Shanghai University of Engineering Science, Shanghai 201620, China (E-mail: guohui@sues.edu.cn)

[c] Associate Professor, School of Mechanical and Automotive Engineering, Shanghai University of Engineering Science, Shanghai 201620, China (E-mail: maminghui@sues.edu.cn)

[d] Ph.D. Shanghai Intelligent and Connected Vehicle R&D Center Co. Ltd, Shanghai 201499, China (E-mail: wu.yuepeng@icv-ip.com)

[e] Associate Professor, Business School, University of Shanghai for Science and Technology, Shanghai 200093, China (E-mail: sdliang@usst.edu.cn)

[f] Professor, School of Mechanical and Automotive Engineering, Shanghai University of Engineering Science, Shanghai 201620, China (E-mail: wysgcd@sues.edu.cn)

## ABSTRACT

Connected and Autonomous Vehicles (CAVs) technology facilitates the advancement of intelligent transportation. However, intelligent control techniques for mixed traffic flow at signalized intersections involving both CAVs and Human-Driven Vehicles (HDVs) require further investigation into the impact of backward-looking effect. This paper proposes the concept of "$1+n+1$" mixed platoon considering the backward-looking effect, consisting of one leading CAV, $n$ following HDVs, and one trailing CAV. The leading and trailing CAVs collectively guide the movement of intermediate HDVs at intersections, forming an optimal control framework for platoon-based CAVs at signalized intersections. Initially, a linearized dynamic model for the "$1+n+1$" mixed platoon is established and compared with a benchmark model focusing solely on controlling the lead vehicle. Subsequently, constraints are formulated for the optimal control framework, aiming to enhance overall intersection traffic efficiency and fuel economy by directly controlling the leading and trailing CAVs in the platoon. Finally, extensive numerical simulations compare vehicle throughput and fuel consumption at signalized intersections under different mixed platoon control methods, validating that considering both front and backward-looking effects in the mixed platoon control method outperforms traditional methods focusing solely on the lead CAV.

---

[b*] Corresponding author 1. Tel.: + 86 13764589361.
   *E-mail address:* guohui@sues.edu.cn (H. Guo).

[c*] Corresponding author 2. Tel.: + 86 18643154316.
   *E-mail address:* maminghui@sues.edu.cn (M. Ma).

# 1. Introduction

With rapid economic development, the global motor vehicle population continues to increase. While the gradual popularization of automobiles has facilitated people's travel and improved their quality of life, it has also brought about environmental pollution, traffic congestion, safety issues, among others. Particularly, urban road traffic problems are becoming increasingly severe, with traffic efficiency optimization at signalized intersections being a crucial aspect in addressing urban road congestion issues. Existing studies have shown that frequent deceleration and stopping of vehicles approaching signalized intersections are the main causes of traffic congestion [1]. Thus, optimizing vehicle trajectories near signalized intersections is of great importance.

In recent years, emerging technologies and theories such as autonomous driving, vehicular networking, and big data have provided new insights and possibilities for addressing issues at urban road traffic signalized intersections. Among these, the operation mode of vehicle platoons is regarded by experts and scholars as a significant direction for the future development of autonomous driving. Existing research has demonstrated that the operation mode of vehicle platoons can effectively alleviate traffic pressure and achieve overall optimization of the traffic system [2]. However, the application scope of fully autonomous vehicle platoon control schemes is currently limited. In real life, it may take decades to convert all manually driven vehicles in the current traffic system into connected autonomous vehicles [3]. It can be foreseen that in the future road network, a complex situation of mixed driving of connected autonomous vehicles and manually driven vehicles will persist for a long time, thus highlighting the importance of research on mixed traffic flow.

Due to the significant potential of mixed traffic flow platoon control in addressing future traffic issues at signalized intersections, an increasing number of experts and scholars are attempting to improve traffic efficiency by controlling mixed traffic flow near signalized intersections.

Although many experts and scholars have conducted extensive research on mixed traffic flow, signalized intersections, or their combination, a systematic and comprehensive theoretical system has not yet been formed. Some research still exhibits certain omissions, and almost all of the following models for manually driven vehicles in mixed traffic flow only apply relatively simple and traditional following models that consider only the preceding vehicle [4], [5], [6],while rarely employing following models that consider the backward-looking effect of vehicles, which is more in line with human driving habits. Therefore, the control frameworks or algorithms obtained often are optimal only under specific conditions.

This paper primarily investigates the scenario of coexistence between HDVs and CAVs at signalized intersections, aiming to improve the performance of the entire mixed traffic flow through the intersection by directly controlling CAVs. Building upon the consideration of the backward-looking effect of vehicles, this paper disassembles the traffic flow into a "1+n+1" mixed platoon microstructure. This structure comprises one leading CAV, $n$ following HDVs, and one trailing CAV. The leading and trailing CAVs collectively guide the movement of intermediate HDVs, proposing an optimal control framework based on platoon for CAVs at signalized intersections. Specifically, the main innovations and contributions of this study are summarized as follows:

（1） Under the premise of considering the backward-looking effect of vehicles, a concept of "1+n+1" mixed platoon is proposed and analyzed for trajectory optimization of autonomous vehicles at signalized intersections in a mixed traffic flow environment;

（2）An optimal control framework for the "1+n+1" mixed platoon considering the backward-looking effect of vehicles is established. The control objective of this framework is to enhance the traffic efficiency at signalized intersections while minimizing fuel consumption during vehicle travel to the intersection, thereby improving the overall performance at signalized intersections under mixed traffic flow conditions;

（3）Extensive simulation experiments validate that compared to traditional control methods focusing solely on the preceding vehicles [34], the proposed optimal control framework considering the backward-looking effect can effectively enhance traffic efficiency near signalized intersections while reducing fuel consumption. Additionally, the impact of different forward and backward-looking influence factors (denoted as p) on the effectiveness of the proposed optimal control framework is compared. The results indicate the universality of the proposed optimal control framework, which can serve as a reference for subsequent research.

The remaining organization of this paper is as follows: Section 2 provides a brief overview of the research status and existing problems in this field. Section 3 demonstrates the specific research scenario and establishes the "1+n+1" mixed platoon model. Section 4 presents the corresponding optimal control framework for the previously proposed mixed platoon model. In Section 5, simulation experiments are conducted to compare and validate the models and control frameworks proposed in this paper. Finally, Section 6 concludes the paper.

## 2. Literature review

2.1 Issues and solutions at signalized intersections

Research on signalized intersections dates back to the 1990s when experts and scholars began evaluating the traffic safety and efficiency near these intersections [7], [8], [9]. It was soon identified that the frequent driving and stopping behavior of individual vehicles approaching intersections are the primary causes of traffic congestion and casualties [10]. To address these issues, various attempts have been made. Cunto et al. proposed a systematic procedure for calibrating and validating microsimulation models for safety performance evaluation, providing an objective and effective method for assessing the safety performance of signalized intersections [11]. Michler et al. studied a driver assistance system for improving traffic efficiency at signalized intersections using predictive traffic state estimation [12]. With the emergence of vehicular networking technology, the approaches to addressing safety and traffic efficiency issues at signalized intersections have diversified. Intelligent signal timing adjustment has been considered to influence the safety and efficiency of signalized intersections [13], [14]. With the continuous development of vehicular networking technology, connected autonomous vehicles can optimize their approach trajectory to intersections in real-time based on accurate information about surrounding traffic participants and traffic signal phases, thereby achieving higher traffic efficiency and lower fuel consumption. Thus, scholars have proposed controlling intersection efficiency directly through controlling connected autonomous vehicles and have provided various control methods, including model predictive control or optimal control [15],

[16], [17], [18], [19], [20]. Additionally, research on cooperative control of connected autonomous vehicles has been prioritized in recent years [21], [22].

2.2 Mixed traffic flow control

Research on mixed traffic flow consisting of connected autonomous vehicles and manually driven vehicles has gradually emerged in recent years. Yao et al. described the following behavior between HDVs and CAVs using different following models, proposing an analysis method for the stability of mixed traffic flow and deriving basic diagrams of mixed traffic flow under different CAV penetration rates, analyzing the influencing factors of basic diagrams [23]. Ghiasi et al. proposed an analytical capability model based on a Markov chain to represent the heterogeneous and stochastic headway distance spatial distribution of mixed traffic flow on highways, which allows the examination of the impact of different CAV technology schemes on mixed traffic capacity and serves as a useful and simple decision-making tool for CAV lane management [24]. Dhatbale et al. applied deep learning technology to mixed traffic flow and used it to extract vehicle motion trajectories [25].

Regarding signalized intersections under mixed traffic flow, scholars have also conducted various discussions. Priemer et al. conducted research on traffic state estimation and traffic signal optimization at signalized intersections under mixed traffic flow in 2009 [26]. Many subsequent scholars have conducted similar studies [27], [28], [29]. However, the above-mentioned studies do not fully cover the research content of signalized intersections under mixed traffic flow. For example, the vehicle control and trajectory optimization problems of connected autonomous vehicles have not been fully discussed. Some scholars have conducted related research. Du et al. used a hierarchical control framework to coordinate a group of connected autonomous vehicles passing through multiple signalized intersections [30]. Jiang et al. proposed an ecological driving system for connected autonomous vehicles at signalized intersections, which optimizes traffic efficiency by optimizing the velocity curve of connected autonomous vehicles [31]. It is evident that these studies focus only on improving the performance of connected autonomous vehicles in mixed traffic flow, neglecting the overall optimization of mixed traffic flow. Subsequently, scholars recognized this issue and conducted related research, attempting to improve intersection traffic performance by constructing mixed platoons and achieved some results, but without providing a clear definition of mixed platoons or exploring their essence [32], [33]. Only recently, Chen et al. provided a clear definition of mixed platoons under certain specific conditions and proposed an optimal control framework for signalized intersections applicable in this scenario [34].

2.3 Vehicle platoon model considering backward-looking effect

In the application of following models for human-Driven vehicles, the aforementioned research only applied following models considering the influence of preceding vehicles, such as the classical optimal velocity model (OVM) and intelligent driver model (IDM). Consequently, the corresponding optimal control frameworks were limited to this consideration. However, in real-life driving, especially in complex traffic environments such as passing through signalized intersections, people often pay attention to vehicles behind them, not just those in front of them. The first to discover and provide a clear answer to this issue was Nakayama et al., who proposed a specific following model considering the backward-looking Effect (BLOV), which focuses on both preceding and following vehicles [35]. Subsequently, many studies have improved, refined, and enhanced based on the BLOV model. For example, Sun et al. proposed the backward-looking velocity difference model (BLVD), which assigns explicit weights to front and backward-looking effects (implicit in the BLOV model) and considers velocity differences [36]. Chen et al. introduced the influence of driver sensory memory on vehicle driving based on the consideration of the backward-looking effect, further expanding micro-vehicle models [37]. Ma et al. first proposed an extended following model considering the change of headway distance with memory based on the BLVD model. Subsequently, combined with connected autonomous vehicles and considering both backward-looking effect and multi-vehicle motion information, an improved all-connected autonomous vehicle following model was proposed [38], [39].

## 3. Problem statement

### 3.1 Scenario setup

This study investigates a typical signalized intersection scenario under mixed traffic environment, where HDVs and CAVs coexist; as shown in Figure 1. Traffic signal lights are deployed at the center to guide vehicles through the intersection. It's noteworthy that autonomous vehicles follow instructions from the Central Cloud Coordinator, which collects information from all relevant vehicles around the intersection and computes the optimal velocity trajectory for each autonomous vehicle. Section 4 discusses the design of the Central Cloud Coordinator control strategy.

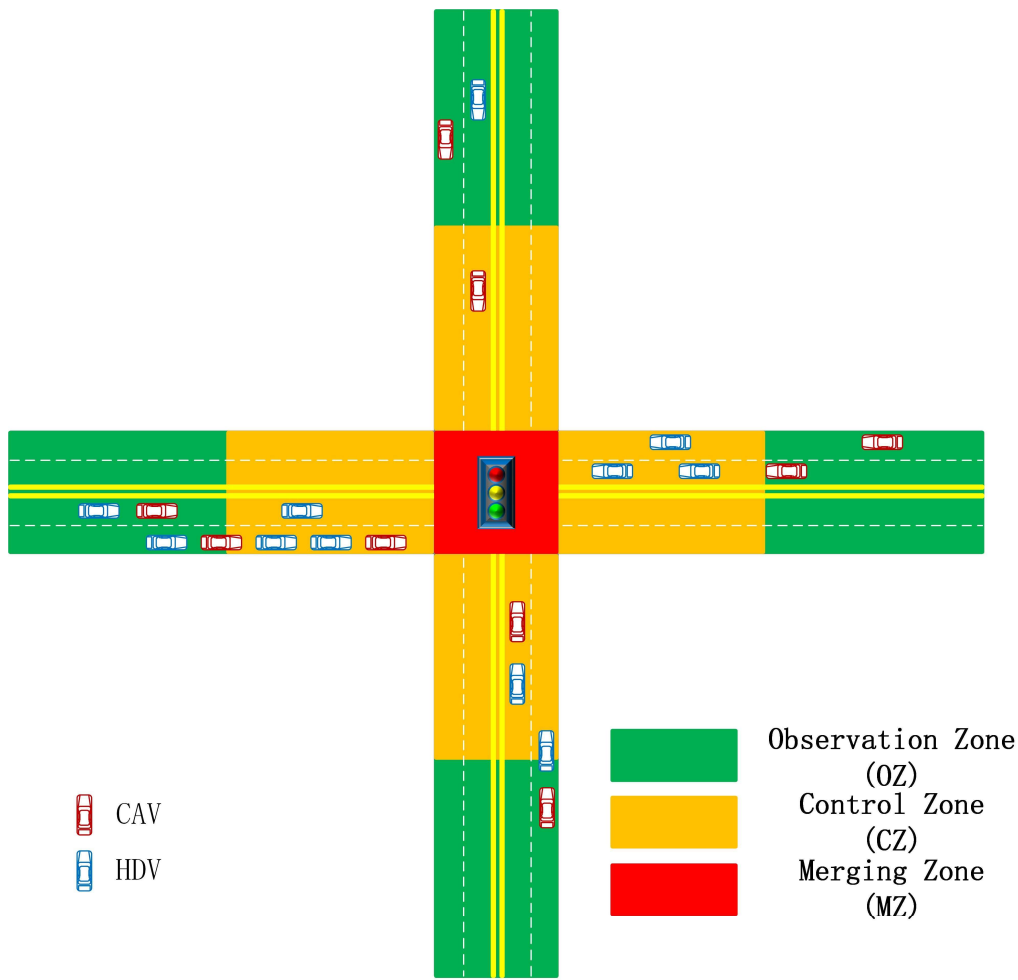

Fig. 1. Illustration of a signalized intersection with mixed traffic flow. Red vehicles represent Connected Autonomous Vehicles (CAVs), capable of sending and receiving vehicle information, and fully autonomous. Blue vehicles represent Human-Driven Vehicles (HDVs), which can only transmit self-vehicle information to other vehicles and are controlled by a car-following model.

Inspired by previous studies on signalized intersections, this paper divides the intersection into three zones, as depicted in Figure 1. The central red square area is referred to as the Merge Zone (MZ), where lateral collisions between relevant vehicles may occur. The yellow road area near the center is the Control Zone (CZ), where CAVs are directly controlled by the Central Cloud Coordinator. The outermost green road area is the Observation Zone (OZ), where the motion behavior of any vehicle is generally unrestricted. The specific ranges of each zone will be discussed in Section 4.

### 3.2 Basic assumptions

Similar to existing research, the following assumptions are made for the ease of control design at signalized intersections, as well as system modeling and dynamic analysis:
(1) All vehicles are connected vehicles, meaning both CAVs and HDVs can transmit their velocity and position to the Central Cloud Coordinator through wireless communication, such as V2I communication, under ideal communication conditions without communication delays or packet loss.
(2) All autonomous vehicles, upon entering the CZ, follow the velocity trajectory assigned by the Central Cloud Coordinator for fully autonomous driving. As for HDVs, they are controlled by human drivers, and we assume a generic car-following model to describe their driving behavior (see Section 3.3 for details).
(3) Lane changing is not allowed in the CZ. Accidental lane-changing behavior may reduce traffic efficiency, especially near intersections. Therefore, lane-changing is only permitted in the OZ, whereas in the CZ, we only need to focus on the longitudinal behavior of each vehicle.
(4) Only the state of a single mixed platoon passing through the signalized intersection is considered, hence the need to address longitudinal collisions between platoons when multiple mixed platoons are operating within the CZ is not required.

### 3.3 Model establishment

#### 3.3.1 Benchmark model

The benchmark model referred to in this paper is the "1+$n$" mixed platoon model proposed by Chen et al. [34], as shown in Figure 2. It consists of a leading CAV and $n$ following HDVs. In the entire mixed traffic flow, each CAV can be designed as the leader of the mixed platoon to guide the motion of the $n$ following HDVs, thereby improving the performance of the entire mixed platoon when passing through the intersection.

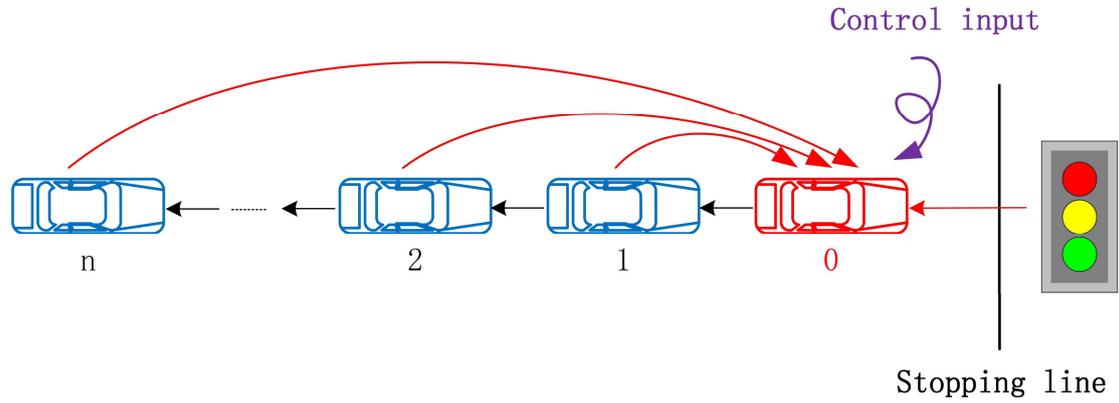

Fig. 2. Illustration of the "1+$n$" mixed platoon. The solid red arrows represent the information flow of the leading CAV (red), which collects information from all subsequent vehicles in the mixed platoon and traffic lights, and has external control input; the black arrows represent the information flow of HDVs (blue), which are controlled by human drivers and only receive information from adjacent leading vehicles.

However, in this benchmark model, only the influence of adjacent leading vehicles is considered for human-driven vehicles (HDVs). In real-life driving scenarios, especially in complex traffic environments such as signalized intersections, drivers tend to pay attention not only to the vehicles in front but also to those behind. Therefore, by considering the backward-looking effect of vehicles, we have established a new mixed platoon model.

### 3.3.2 Dynamical modeling of the "1+n+1" mixed platoon system

This paper proposes the concept of the "1+$n$+1" mixed platoon, as shown in Figure 3. This model consists of a leading CAV, $n$ following HDVs, and a trailing CAV. Similarly, in the entire mixed traffic flow, each CAV can be designed as the leader or the trailing vehicle of the mixed platoon. Compared to the benchmark model, since the HDVs in this model consider the influence of vehicles in both the front and rear directions, the leading and trailing CAVs can jointly guide the motion of the $n$ HDVs in the middle, thereby improving the performance of the entire mixed platoon when passing through the intersection.

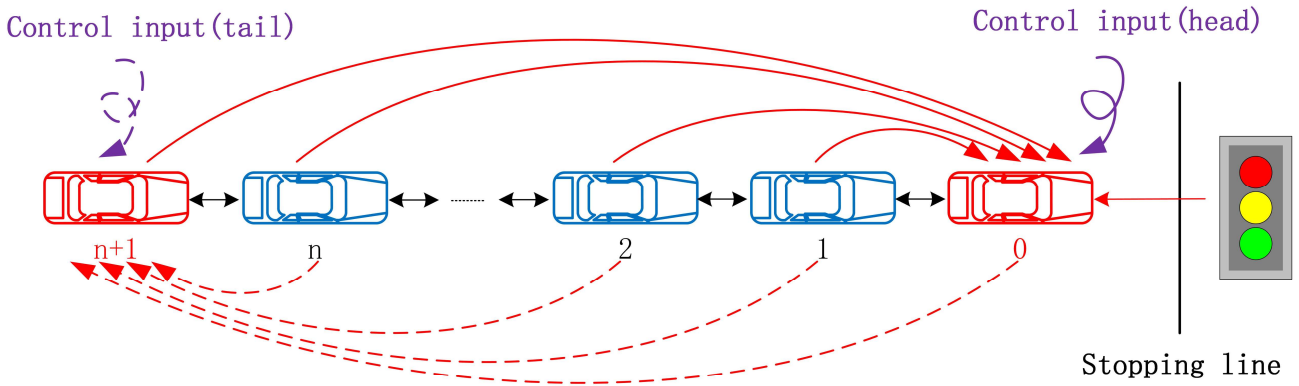

Fig. 3. Illustration of the "1+$n$+1" mixed platoon. The solid red arrows represent the information flow of the leading CAV (red), which collects information from all subsequent vehicles in the mixed platoon and traffic lights, and has external control input; the red dashed

arrows represent the information flow of the trailing CAV (red), which collects information from all vehicles in front of the mixed platoon, and has external control input; the black arrows represent the information flow of HDVs (blue), which are controlled by human drivers, consider the backward-looking effect but also only receive information from adjacent vehicles.

In existing research, many efforts have been made to describe the car-following dynamics of HDVs and several important models have been developed, such as OVM and IDM. However, these models only consider the interaction of information between the current vehicle and its adjacent leading vehicle, ignoring the influence of the trailing vehicle on the current vehicle. Therefore, to better conform to the driving habits of people in reality, car-following models for HDVs considering the backward-looking effect have been studied, such as the BLOV and BLVD models. Most of these models can be interpreted as the following general form:

$$\dot{v}_i(t) = F\left(d_i(t), \dot{d}_i(t), v_i(t)\right) \qquad i \in (1, n) \tag{1}$$

Consider the $n$ HDVs in the mixed platoon shown in Figure 2. We represent the position and velocity of vehicle $i$ at the current time $t$ as $x_i(t)$ and $v_i(t)$ respectively. Then, $d_i(t) = x_{i-1}(t) - x_i(t)$ or $d_i(t) = x_i(t) - x_{i+1}(t)$ represents the headway distance of vehicle $i$ with its adjacent vehicles, and $\dot{d}_i(t) = v_{i-1}(t) - v_i(t)$ or $\dot{d}_i(t) = v_i(t) - v_{i+1}(t)$ represents the relative velocity of vehicle $i$ with its adjacent vehicles. This implies that the acceleration of vehicle $i$ is determined by the headway distance, relative velocity, and its own velocity.

In this paper, we require the mixed platoon to pass through the intersection at a predetermined equilibrium velocity $v^*$. In the equilibrium traffic state, we have $\dot{d}_i(t) = 0, i \in (1, n)$, thus each vehicle in the mixed platoon has a corresponding equilibrium headway distance $d^*$. In this state, (1) transforms into (2):

$$F(d^*, 0, v^*) = 0 \qquad i \in (1, n) \tag{2}$$

To assess $(d^*, v^*)$ in the equilibrium state, we strike a balance between model fidelity and computational tractability while considering the vehicle's backward-looking effect. Eventually, we opt to apply the classic BLOV model as the specific car-following model in the optimal control formulation in Section IV. In the BLOV model, the general expression (1) for the car-following dynamics of HDVs can be represented in the following form:

$$\dot{v}_i = a\{[V_F(x_{i-1} - x_i) + V_B(x_i - x_{i+1})] - v_i\} \tag{3}$$

Here, $V_F(x)$ serves as the forward-looking OV function, representing the expected velocity-headway distance function between the current vehicle and the preceding one, while $V_B(x)$ is the backward-looking OV function, representing the expected velocity-headway distance function between the current vehicle and the following one.

We choose two OV functions as

$$V_F(x) = p \frac{v_{max}^F}{2} [\tanh(x - h_c) + \tanh(h_c)] \tag{4}$$

$$V_B(x) = -(1-p) \frac{v_{max}^B}{2} [\tanh(x - h_c) + \tanh(h_c)] \tag{5}$$

In Equation (4) and (5), $h_c$ denotes the safety distance for the smooth operation of the vehicle platoon, $v_{max}^F$ and $v_{max}^B$ respectively represent the maximum velocity in the forward and backward directions, and $p$ is the proportional coefficient for the weighting of forward and backward considerations, Abbreviations: forward and backward-looking influence factors.

For the leading CAV, indexed as vehicle 0, its acceleration signal is utilized as the control input

$u(t)$ for the lead vehicle. Thus, the longitudinal dynamics of the lead vehicle can be expressed in the following second-order form:

$$\begin{cases} \dot{x}_0(t) = v_0(t) \\ \dot{v}_0(t) = u_0(t) \end{cases} \quad (6)$$

Similarly, for the trailing CAV, indexed as vehicle n+1, its acceleration signal is utilized as the control input $u_{n+1}(t)$ for the trailing vehicle. Thus, the longitudinal dynamics of the trailing vehicle can be expressed in the following second-order form:

$$\begin{cases} \dot{x}_{n+1}(t) = v_{n+1}(t) \\ \dot{v}_{n+1}(t) = u_{n+1}(t) \end{cases} \quad (7)$$

It is noteworthy that the acceleration signals of the leading and trailing CAVs are the only external control inputs to the entire mixed platoon system as shown in Fig. 3. By aggregating the states of these two CAVs and the intermediate HDVs, we obtain the complete "1+n+1" mixed platoon system.

## 4 Optimal control framework

After the basic dynamics analysis of the fundamental characteristics of the proposed "1+n+1" mixed platoon, in this section, we proceed to establish the optimal control framework for the mixed platoon at signalized intersections.

### 4.1 Cost function

In the optimal control framework for the "1+n+1" mixed platoon at the signal intersection, the main control objectives are as follows: Firstly, to ensure that the leading CAV arrives at the intersection stop line when the traffic signal turns green. Secondly, to control the trailing CAV to approach the preceding vehicle as closely as possible, ensuring safety and compact spacing between vehicles in the mixed platoon to enhance traffic efficiency when passing through the intersection. Additionally, to stabilize the HDVs in the middle at the desired equilibrium velocity $v^*$. Furthermore, we also aim at minimizing the fuel consumption of the entire mixed platoon as it approaches the intersection. Therefore, the cost function is defined as:

$$J = \varphi(X(t_f)) + \int_{t_0}^{t_f} L(X(t), u(t)) dt \quad (8)$$

where $t_0$ denotes the time when the leading CAV is about to enter CZ, i.e., the moment it reaches the boundary between CZ and OZ, and $t_f$ represents the time when the leading CAV is about to enter MZ, i.e., the moment it reaches the stop line. The specific selection of $t_f$ will be discussed in detail in Section 4.4.

As the terminal cost function in (8), the variable definition $\varphi(X(t_f))$ represents the deviation between the final state of the system and the desired state. It is defined as:

$$\varphi(X(t_f)) = \omega_1(x_0(t_f) - x_{star})^2 + \omega_2 \sum_{i=0}^{n+1}(v_i(t_f) - v^*)^2 + \omega_3 (x_0(t_f) - x_{n+1}(t_f))^2 \quad (9)$$

In Equation (9), $\omega_1$、$\omega_2$、$\omega_3$ are penalty weight coefficients for the position deviation of the leading CAV, the velocity deviation of all vehicles in the mixed platoon, and the position gap deviation between the leading and trailing CAV of the mixed platoon, respectively. $x_0(t_f)$ represents the actual position of the leading CAV at $t=t_f$, and $x_{n+1}(t_f)$ represents the actual position of the trailing CAV at $t=t_f$. $x_{star}$ denotes the final target position of the leading CAV, i.e., the location of the intersection stop line. The specific selection of the ideal equilibrium velocity $v^*$ for CAV and the target position $x_{star}$ will be discussed separately in Sections 4.2 and 4.3.

In Equation (8), $L(X(t), u(t))$ represents the transient fuel consumption of the mixed platoon at time $t$, defined as:

$$L(X(t), u(t)) = G_0(t) + \sum_{i=1}^{n} G_i(t) + G_{n+1}(t) \tag{10}$$

Here, $G_0(t)$ 和 $G_i(t)$, $(i = 1, \cdots, n)$, $G_{n+1}(t)$ represent the transient fuel consumption of the leading CAV, the following HDVs, and the trailing CAV, respectively. We employ Akcelik's fuel consumption model as a specific model to calculate transient fuel consumption [40].

$$G_i(t) = \alpha + \beta_1 P_T(t) + (\beta_2 m a_i(t)^2 v_i(t))_{a_i(t)>0} \tag{11}$$

Here, $m$ represents the vehicle mass, $(\beta_2 m a_i(t)^2 v_i(t))_{a_i(t)>0}$ denotes the additional inertial (engine/internal) resistance power when the vehicle accelerates. Here, $\alpha$ stands for the idle fuel consumption rate, and $P_T$ represents the total power driving the vehicle, which includes engine drag power, rotational inertia, air friction, and other energy losses; it can be calculated as follows:

$$P_T(t) = \max\{0, d_1 v_i(t) + d_2 v_i(t)^2 + d_3 v_i(t)^3 + m a_i(t) v_i(t)\} \tag{12}$$

As suggested by Akcelik, we consider typical settings for the parameter values in Akcelik's fuel consumption model (11) and (12). For specific data, please refer to Table 1.

## 4.2 Terminal velocity

In this section, we will discuss how to design the desired equilibrium velocity $v^*$, which also represents the terminal velocity in the terminal cost function (9). Existing research has mainly focused on the individual control of autonomous vehicles, and to improve intersection efficiency, the terminal velocity of autonomous vehicles is typically set to the maximum velocity limit; see Asadi and Vahidi [17] and Jiang et al. [20]. However, considering the presence of other HDVs at intersections, such settings in the aforementioned works may not be the optimal choice for the entire mixed intersection. Chen et al. [34] provide us with a new approach to design the desired equilibrium velocity.

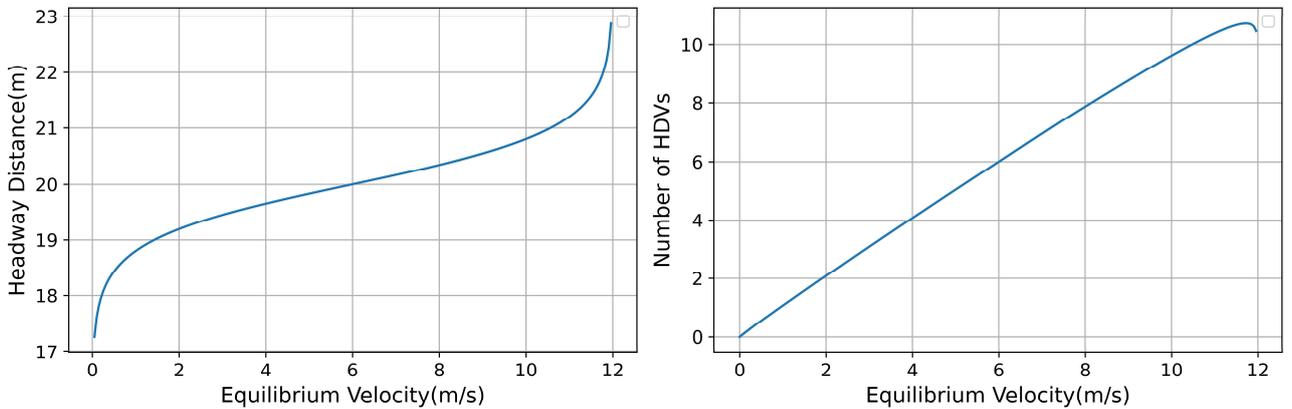

Fig.4. The relationship between the equilibrium velocity of the mixed platoon (i.e., terminal velocity) and the number of HDVs passing

through, as BLOV model is adopted as the car-following model with parameters shown in Table 1. As the equilibrium velocity increases, the equilibrium headway distance between vehicles increases, and the number of vehicle passages within a certain green light time first rises and then falls. Therefore, for a constant flow rate, there exists a maximum equilibrium velocity value that maximizes the number of HDVs passing through, as shown by the peak in the right graph.

When designing the terminal velocity, our objective is to maximize the number of vehicles passing through the intersection in a certain green light phase duration $T_{gre}$ . Taking a "1+n+1" mixed platoon as an example, from the equilibrium (2) of the HDV-following model, it is evident that the equilibrium headway distance $d^*$ depends on the equilibrium velocity $v^*$. For a constant green light phase duration, our optimization goal is to maximize the number $n$ of HDVs passing through during $T_{green}$. Hence, we can obtain the following result.

The optimal velocity $v^*$ can be obtained by solving the following optimization problem:

$$\text{argmax}_{v^*} \quad n = \frac{v^* T_{green}}{d^*} \tag{13}$$

subject to : $\quad F(d^*, 0, v^*) = 0$

Recalling that we derived an explicit expression $F(\cdot)$ for the HDV-following model using the BLOV model (3). Inspired by Nakayama and Sugiyama [35] and Ma et al. [38], we set the relevant parameters as follows, based on typical parameter settings for the BLOV model, as shown in Table 1. When the traffic light turns green, the leading CAV is expected to reach the stop line. Therefore, if the mixed platoon is in equilibrium at this time, the CAVs at both ends and the HDVs in the middle will travel at the same velocity, i.e., $F(\cdot) = \dot{v}_i = 0$, which applies simultaneously to any vehicle in the mixed platoon in equilibrium. Thus, we obtain the relationship between velocity and headway distance at equilibrium based on the BLOV model as follows.

$$d_i = arctanh\left(\frac{v_i}{2p-1*0.5v_{max}} - \tanh(hc)\right) + hc \tag{14}$$

From Figure 4, it can be observed that at equilibrium, the equilibrium headway distance of vehicles is typically a monotonically increasing function relative to the equilibrium velocity. There also exists a maximum number of vehicles corresponding to the optimal equilibrium velocity $v^*$ and equilibrium headway distance $d^*$. The optimal terminal velocity $v^*$ can be obtained by solving (13).

## 4.3 Constraints

To implement the designed CAV controller in practice, several constraints need to be considered, including process constraints and terminal constraints.

In terms of process constraints, the first concern is the safety constraint for the entire mixed platoon, ensuring that each vehicle in the mixed platoon maintains a minimum safe distance $d_{safe}$ from the preceding vehicle. Here, $L_{veh}$ represents the length of the vehicle.

$$x_{i-1}(t) - x_i(t) - L_{veh} \geq d_{safe}, \text{for } t_0 \leq t \leq t_f, i = 1,2,\cdots,n+1 \tag{15}$$

Next is the effectiveness constraint for the trailing CAV in the mixed platoon. To ensure that the trailing CAV effectively closes in on the preceding vehicle within the entire control zone, its distance from the front vehicle must be maintained within a certain range, specifically between the minimum safe distance $d_{safe}$ and the safe gap for smooth vehicle operation $h_c$.

$$d_{safe} + L_{veh} \leq x_{i-1}(t) - x_i(t) \leq h_c, \text{for } t_0 \leq t \leq t_f, i = n+1 \tag{16}$$

Finally, there are practical constraints on the velocity and acceleration values of each vehicle in the mixed platoon. $v_{max}$ represents the maximum velocity, while $a_{min}$ and $a_{max}$ represent the minimum and maximum accelerations, respectively. Therefore, they must satisfy the following:

$$0 \leq v_i(t) \leq v_{max}, \text{for } t_0 \leq t \leq t_f, i = 0,1,2,\cdots,n+1 \qquad (17)$$

$$a_{min} \leq a_i(t) \leq a_{max}, \text{for } t_0 \leq t \leq t_f, i = 0,1,2,\cdots,n+1 \qquad (18)$$

For terminal constraints, our main focus is on the terminal positions of the leading and trailing CAVs. Recall that the deviation of the terminal position $x_0(t_f)$ of the leading CAV from the target position $x_{star}$ has already been accounted for in the terminal cost function (9). As for the trailing CAV, there is no fixed terminal position $x_{n+1}(t_f)$ relative to a predefined target position. The control objective for the trailing CAV has been clearly expressed in the terminal cost function (9), and constraints on its terminal position are already included in the process constraints (16). Therefore, here, we only need to add an inequality constraint, requiring that the leading CAV neither overshoots the stop line nor maintains a large distance from it, as follows:

$$0 \leq x_0(t_f) \leq x_0^{max}(t_f) \qquad (19)$$

where $x_0^{max}$ represents the maximum tolerance distance of the leading CAV from the stop line. In the control of mixed traffic flow at signalized intersections, if the terminal time $t_f$ and the corresponding terminal position $x_0(t_f)$ are fixed in advance, similar to single CAV control algorithms, the feasible region of control inputs for the leading CAV in this case would be greatly restricted, potentially compromising the optimal performance of the entire mixed platoon system.

## 4.4 Optimal control formulation

Bringing together the design of the above cost function and the constraint conditions, the overall optimal control problem can be formulated as:

$$argmin_{u_0(t),u_{n+1}(t)} J = \varphi\left(X(t_f)\right) + \int_{t_0}^{t_f} L(X(t), u(t))dt \qquad (20)$$

subject to : (3),(4),(5),(15),(16),(17),(18),(19).

Before solving problem (20), it is necessary to first compute the optimal terminal velocity $v^*$ by solving problem (13). Additionally, the terminal time $t_f$ needs to be determined beforehand, which is the subject of our discussion next.

Regarding the determination of the terminal time $t_f$, we first refer to the approach proposed by Asadi and Vahidi [17]. This paper suggested a practical method for selecting the target green phase window after obtaining the traffic phase diagram in advance using V2I technology, as follows:

$$[v_{low}, v_{high}] = \left[\frac{D_k}{r_j - t}, \frac{D_k}{g_j - t}\right] \cap [v_{min}, v_{max}] \qquad (21)$$

In Equation (21), $D_k$ represents the distance from CAV $k$ to the stop line; $t$ denotes the current time; $r_j$ is the start time of the next red light phase; $g_j$ is the start time of the next green light phase; $v_{min}$ and $v_{max}$ are the velocity limits for CAVs. Considering the green light time from (13), we have $T_{green} = r_j - g_j$, where $T_{green} > 0$. The non-empty intersection $[v_{lo}, v_{high}]$ in (21) represents the feasible velocity window that allows CAVs to pass through the intersection without idling. Then, to maximize the number of vehicles passing through within a certain green phase, CAVs obtain the target velocity $v_{target} = v_{high}$, and subsequently, the corresponding terminal time $t_f = \frac{D_k}{v_{target}}$ can be calculated.

However, in our study, there exists an optimal terminal velocity as designed in Section 4.2. If the target velocity is set to the maximum velocity $v_{max}$, then optimizing the velocity trajectory becomes meaningless. Therefore, we also refer to Chen et al. [34] and optimize (21), yielding the following:

$$[v_{low}, v_{high}] = \left[\frac{D_k}{r_j-t}, \frac{D_k}{g_j-t}\right] \cap \left[v_{min}, \frac{v_{max}+v^*}{2}\right] \quad (22)$$

Where $v^*$ represents the optimal velocity calculated from (13). Similar to (21), for the non-empty intersection $[v_{low}, v_{high}]$, the target velocity $v_{target} = v_{high}$ is chosen. The corresponding terminal time is then calculated as $t_f = \frac{D_k}{v_{target}}$. With this approach, the terminal time in the cost function (8) of the optimal control formulation is determined.

In summary, the optimal control formulation (20) is now fully defined. To solve problem (20) numerically, as it is a high-order nonlinear optimal control problem, we can use the pseudo-spectral method to transform it into a nonlinear programming (NLP) problem [41]. Alternatively, several practical software toolkits can be utilized to solve this problem directly; for example, the PYSWARM (Python Optimization) toolkit.

## 5 Simulation results and discussion

In this section, we conducted large-scale traffic simulation experiments, which involved implementing and evaluating the optimal control framework proposed in Section 4 considering the backward-looking effect, using particle swarm optimization algorithm. We compared this framework with scenarios where no control was applied to the vehicle platoon at the signalized intersection and with the optimal control framework of the "1+$n$" mixed platoon proposed by previous researchers, which only considered adjacent leading vehicles.

### 5.1 Simulation environment and parameters

In the experiments conducted in this paper, we utilized Python version 3.10.9 to establish the simulation environment, which involved creating a virtual signalized intersection scenario. During each simulation run, a mixed platoon was generated at the beginning of each loop. We assumed that the leading and trailing vehicles in each mixed platoon were CAVs, while the remaining vehicles were randomly generated as CAVs or HDVs based on the CAV penetration rate. The initial velocity and headway distance of each vehicle in the mixed platoon were randomly generated while ensuring safety. Subsequently, we determined the simulation timestep by balancing the total simulation duration and fidelity. Specific simulation parameters are provided in Table 1. The simulations were executed on an Intel Core i7-12700K processor.

Table1

Parameters in the simulation

| Type | Parameters(unit) | Values |
| --- | --- | --- |
| Vehicle | $v_{min}(m/s)$ | 0 |
| | $v_{max}, v_{max}^F, v_{max}^B (m/s)$ | 15 |
| | $a_{max}(m/s^2)$ | 3 |
| | $a_{min}(m/s^2)$ | -6 |
| | $h_c(m)$ | 20 |
| | $p$ | 0.9 |
| | $a(s^{-1})$ | 0.85 |
| | $\omega_1$ | $10^5$ |
| | $\omega_2$ | $10^4$ |
| | $\omega_3$ | $10^2$ |
| | $L_{veh}(m)$ | 5 |
| | $\alpha(mL/s)$ | 0.666 |
| | $\beta_1(mL/s)$ | 0.072 |
| | $\beta_2(mL/(kJ \cdot m/s^2))$ | 0.0344 |
| | $d_1(kN)$ | 0.0269 |
| | $d_2(kN/(m/s))$ | 0.0171 |
| | $d_3(kN/(m/s)^2)$ | 0.000672 |
| | $m(kg)$ | 1680 |
| Infrastructure | Control zone length (m) | 300 |
| | Observation zone length (m) | 500 |
| Simulation | $\Delta t(s)$ | 0.5 |

## 5.2 Validity analysis

This paper primarily validates the effectiveness of the proposed optimal control framework by comparing and analyzing the average time taken by vehicles to pass through the control zone, the corresponding average headway distance, and the actual maximum number of vehicles passing through the intersection within the same green phase. Three scenarios were examined: (a) a mixed platoon with no control, (b) a mixed platoon with only the leading CAV controlled, and (c) a mixed platoon with both the leading and trailing CAVs controlled simultaneously.

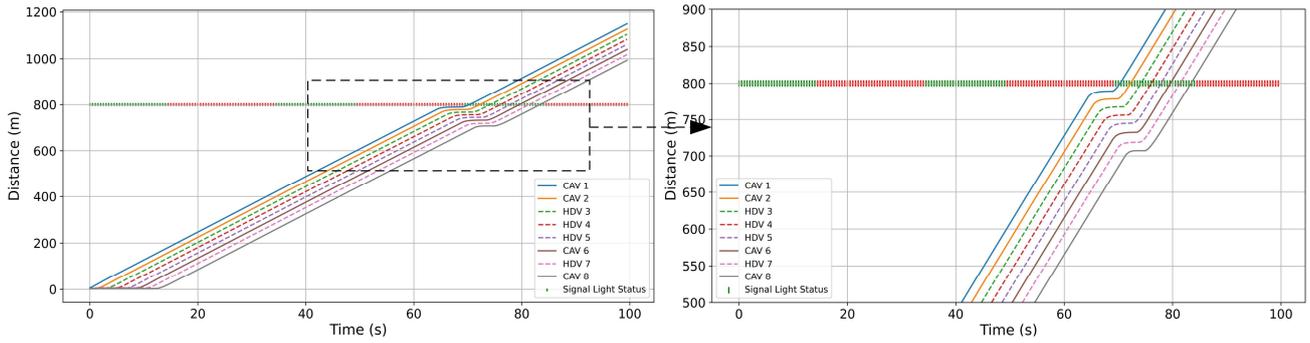

(a) No control input assigned to CAVs

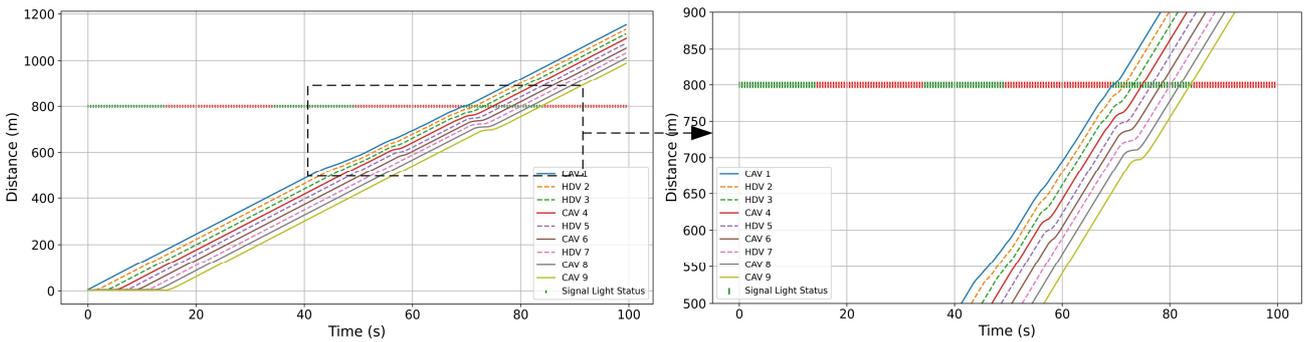

(b) 1+n mixed platoon control input assigned to CAVs

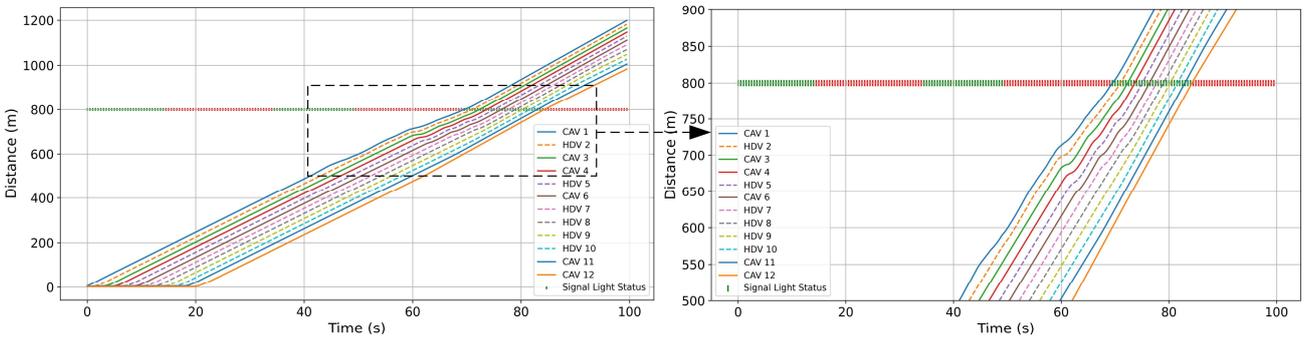

(c) 1+n+1 mixed platoon control input assigned to CAVs

Fig.5. For the scenario with MPR = 50%, trajectory plots were generated for three cases within the same traffic light cycle: no control, "1+n" mixed platoon control, and "1+n+1" mixed platoon control. Each set of plots includes an overall view of the entire simulation process as well as detailed zoomed-in views of the control zone. Solid lines represent the trajectories of CAVs, while dashed lines represent the trajectories of HDVs.

### Table2

Traffic efficiency comparison of the three control methods

|  | No control | "1+n" control | "1+n+1" control |
|---|---|---|---|
| Average travel time(s) | 29.07 | 27.12 | 24.78 |
| Average headway distance(m) | 21.51 | 18.75 | 17.56 |
| Idling time (s) | 22.60 | 0 | 0 |

Here, the paper simulated the scenario with a CAV Market Penetration Rate (MPR) of 50%, which corresponds to distributing nearly equal numbers of CAVs and HDVs randomly in each mixed platoon for the three aforementioned scenarios. In Figure 5, three sets of vehicle trajectories are depicted, representing a mixed platoon with no control input (a), a mixed platoon with only the lead CAV controlled (b), and a mixed platoon with both the lead and tail CAVs controlled (c). It is evident from Figure 5 that the initial states of the mixed platoons in all three scenarios are identical.

In Figure 5 (a), due to the absence of any control input for the first vehicle in the mixed platoon, it moves freely towards the stop line at the signalized intersection when the traffic signal is red, causing subsequent vehicles to queue up due to the red light, resulting in idling and fuel wastage.

In Figure 5 (b), under "1+n" mixed platoon control, the platoon no longer queues up, and the lead CAV guides the following HDVs through the intersection at appropriate velocity, thereby improving the traffic efficiency and fuel economy of the mixed platoon to some extent when passing through the signalized intersection. However, combining this with the observations from Table 2, it can be noted that due to the lack of tail vehicle control and the absence of consideration for the backward-looking effect, the improvement in traffic efficiency when the mixed platoon passes through the signalized intersection under "1+n" mixed platoon control is not particularly significant, and the headway distance between adjacent vehicles remains slightly large.

In Figure 5 (c), due to the "1+n+1" mixed platoon control method considering the backward-looking effect of vehicles, coupled with the findings from Table 2, it can be observed that the headway distance between adjacent vehicles significantly decreases. The average time for vehicles to pass through the control zone also decreases noticeably. Consequently, the number of vehicles passing through the intersection within the same green light phase significantly increases, demonstrating the substantial effectiveness of the proposed "1+n+1" mixed platoon control method in enhancing traffic efficiency at signalized intersections.

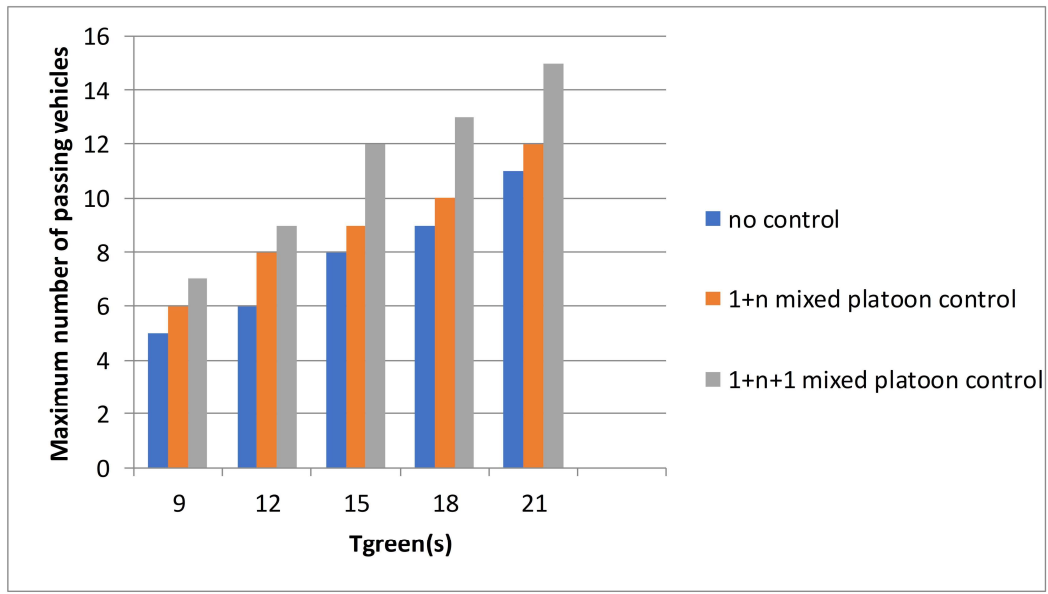

Fig.6. A comparative diagram illustrating the maximum number of vehicles passing through under different green light phase durations for the three control methods. The horizontal axis represents the duration of the green light phase, while the vertical axis represents the maximum number of vehicles passing through.

To further demonstrate the significant improvement in traffic efficiency at signalized intersections brought about by the proposed "1+n+1" mixed platoon control method, we present a schematic diagram illustrating the maximum number of vehicles passing through under different green light phase durations for the three control methods, as shown in Figure 6. It is easy to observe that regardless of the duration of the green light phase, controlling the mixed platoon consistently enhances traffic efficiency at signalized intersections. However, as previously analyzed, under the "1+n" mixed platoon control, due to the lack of consideration for vehicle backward-looking effects within the platoon, the vehicles in the platoon do not follow closely enough, resulting in a less significant improvement in traffic efficiency compared to the uncontrolled state.

In contrast, under the proposed "1+n+1" mixed platoon control method, we can see a relatively significant improvement in traffic efficiency at signalized intersections compared to both the uncontrolled state and the "1+n" mixed platoon control, across various durations of the green light phase. Upon careful analysis, we notice that as the duration of the green light phase increases, the ability of the "1+n+1" mixed platoon control method to improve traffic efficiency slightly decreases. This is because with an increasing number of vehicles passing through per unit time under constant control area length, more time is required for the control applied to the tail CAV to propagate through the entire platoon to achieve a stable state. In such cases, we can adjust the length of the control area appropriately to achieve our control objectives. The specific impact of the control area length on the proposed "1+n+1" mixed platoon control method will be further discussed in Section 5.3.

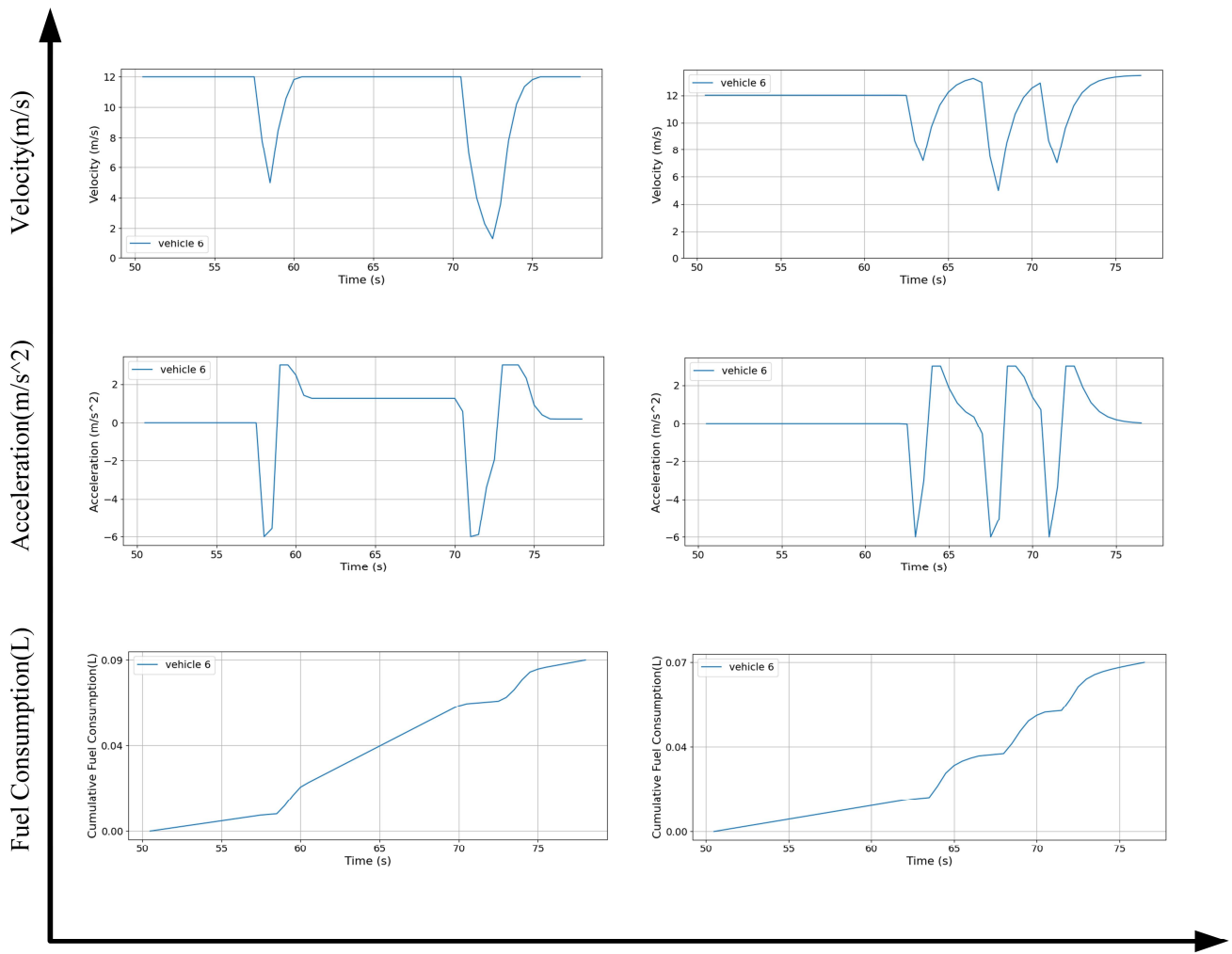

(a) 1+n mixed platoon control　　　　(b) 1+n+1 mixed platoon control

Fig.7. The comparative schematic diagram illustrates the velocity, acceleration, and cumulative fuel consumption of Vehicle 6 under two control methods within the control area.

Table3

Fuel consumption comparison of the two control methods

| Vehicle number | "1+n" control(L) | "1+n+1" control(L) |
| --- | --- | --- |
| 1 | 0.154 | 0.109 |
| 2 | 0.104 | 0.079 |
| 3 | 0.101 | 0.065 |
| 4 | 0.087 | 0.072 |
| 5 | 0.087 | 0.072 |
| 6 | 0.088 | 0.072 |
| 7 | 0.047 | 0.058 |
| 8 | 0.044 | 0.044 |
| 9 | 0.046 | 0.033 |
| 10 | None | 0.033 |
| 11 | None | 0.034 |
| 12 | None | 0.127 |
| Average fuel consumption | 0.084 | 0.066 |

To demonstrate that the "1+n+1" mixed platoon control method proposed in this paper can also improve fuel economy for vehicles, we present the performance of a single vehicle (identified as Vehicle 6) in Figure 7. We compare the variations in velocity, acceleration, and fuel consumption of the same vehicle under two control methods. In Figure 7(a), under "1+n" mixed platoon control, Vehicle 6 does not decelerate to zero, avoiding prolonged engine idling. However, the vehicle's acceleration and deceleration are too large, which can also lead to decreased fuel economy. In contrast, under "1+n+1" mixed platoon control, although the acceleration and deceleration cycles of Vehicle 6 increase, the variation in velocity significantly reduces, resulting in smoother vehicle trajectories and improved fuel economy. Consequently, the entire platoon is also more likely to stabilize. Overall, the accumulated fuel consumption for Vehicle 6 within the control area under "1+n" mixed platoon control is approximately 0.09 liters, whereas under "1+n+1" mixed platoon control, it is only 0.07 liters. Additionally, as shown in Table 3, both the individual vehicle fuel consumption and the average platoon fuel consumption are lower under "1+n+1" mixed platoon control compared to "1+n" mixed platoon control, indicating better fuel economy.

## 5.3 Sensitivity analysis

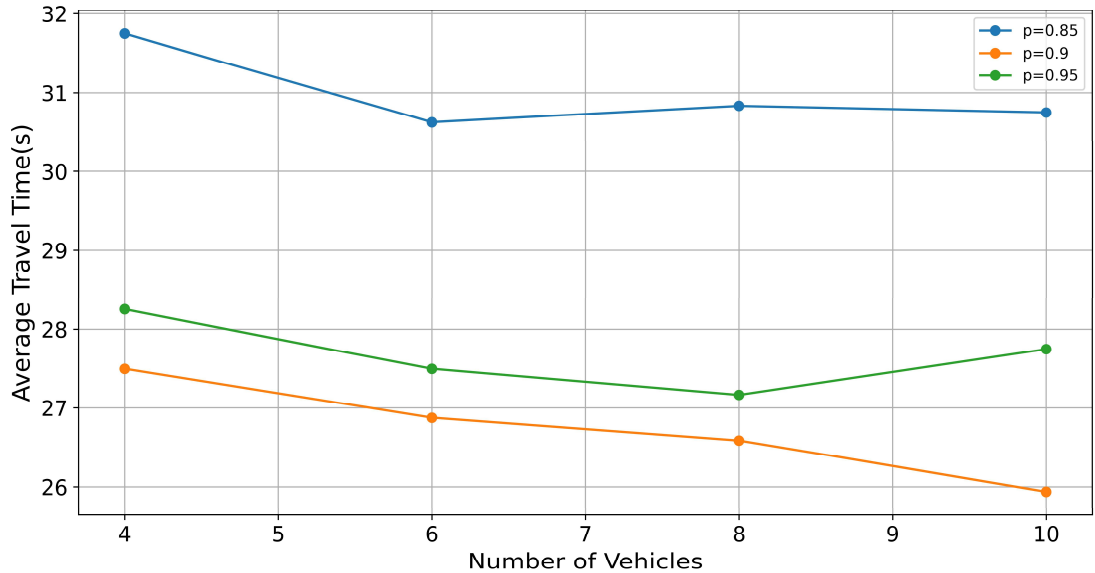

Fig.8. Under "1+n+1" mixed platoon control, the average time for each vehicle to pass through the control area in platoons composed of different numbers of vehicles under different forward and backward-looking influence factors $p$ is illustrated in Figure 8. The horizontal axis represents the number of vehicles in the platoon, while the vertical axis represents the average time for each vehicle in the platoon, excluding the lead and tail vehicles, to pass through the control area.

As shown in Figure 8, we present the average time for mixed platoons composed of different numbers of vehicles to pass through the control area under different forward and backward-looking influence factors $p$. The forward and backward-looking influence factor $p$ is a crucial parameter in the car-following models that consider backward-looking effects. When $p=1$, the car-following model considering backward-looking effects transforms into a conventional model that only considers vehicles in front. Therefore, selecting an appropriate $p$ is a crucial issue that many experts and scholars must address when establishing models considering backward-looking effects.

In this paper, the primary consideration for selecting $p$ is how to maximize the efficiency of mixed vehicle platoons when approaching signalized intersections. Therefore, we present Figure 8, from which we can observe that when $p=0.85$, the average time for the platoon to pass through the control area is relatively long. This is because when we focus too much on the rear vehicles, it tends to reduce the attention to the front vehicles, resulting in a decrease in the overall follow-up density of the platoon, where the platoon's follow-up density refers to the relative velocity and headway distance of each vehicle in the platoon.

However, when $p=0.95$, we find a significant decrease in the time for the platoon to pass through the control area compared to when $p=0.85$. This is because at this time, the vehicles in the platoon pay too much attention to the front vehicle. When there are fewer vehicles, the relative velocity of the vehicles in the platoon are naturally higher. However, we find that when the number of vehicles in the platoon reaches a certain level, the overall average travel time of the platoon will increase. This is because at this time, due to insufficient attention to the rear vehicles in the platoon, the control effect added to the last vehicle in the platoon has a small effect on all vehicles in the front platoon, and it also takes more time for this control effect to be transmitted to the vehicles closer to the front of the platoon. This leads to an increase in the headway distance between vehicles, and the overall follow-up density of the platoon is not too high. It is foreseeable that when the number of vehicles in the platoon is large enough, the effect of controlling the rear vehicles to influence the front platoon will completely fail.

When $p=0.9$, we find that regardless of the number of vehicles in the platoon, the average time for the platoon to pass through the control area is lower than the previous two cases. This is because at this time, the vehicles in the platoon have reached a relatively balanced point in terms of attention to both the front and rear directions. In this case, each vehicle in the platoon maintains a relatively high travel velocity and a relatively small headway distance, resulting in a relatively optimal overall follow-up density of the platoon. It is also noticeable that when $p=0.9$, as the number of vehicles in the platoon increases, the average travel time of the platoon will decrease, which further illustrates that the backward-looking effect of the vehicles in the platoon will not decrease or even fail as the number of vehicles increases. At this point, the platoon has reached a relatively perfect operating state. This also demonstrates the universality of the proposed control framework and can provide some reference for subsequent research.

Furthermore, this study also presents the variations in the average velocity of mixed platoons composed of different number of vehicles under different forward and backward-looking influence factors $p$, as shown in Figure 9. It is observed that when $p=0.9$, regardless of the number of vehicles in the platoon, the magnitude of change in the average travel velocity of the platoon has a significant advantage over the other two groups. At the same time, as the number of vehicles in the platoon increases, the stability of the platoon also gradually improves, once again demonstrating the universality of the optimal control framework proposed in this paper, and adopting $p=0.9$ is a relatively optimal choice.

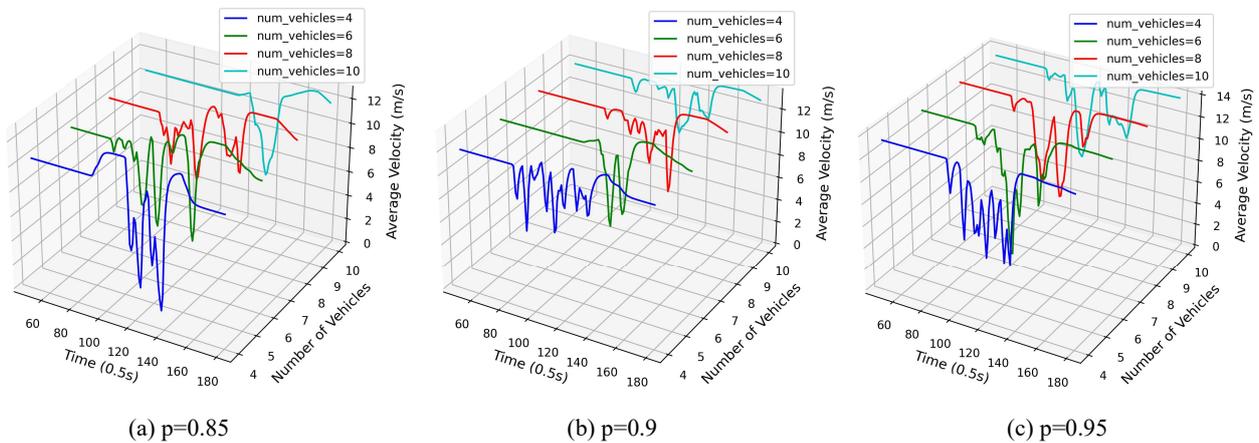

(a) p=0.85　　　　　　　　　　　　　(b) p=0.9　　　　　　　　　　　　　(c) p=0.95

Fig.9. Graph showing the average velocity variation of platoons composed of different number of vehicles under different forward and backward-looking influence factors $p$ in the "1+n+1" mixed platoon control scenario.

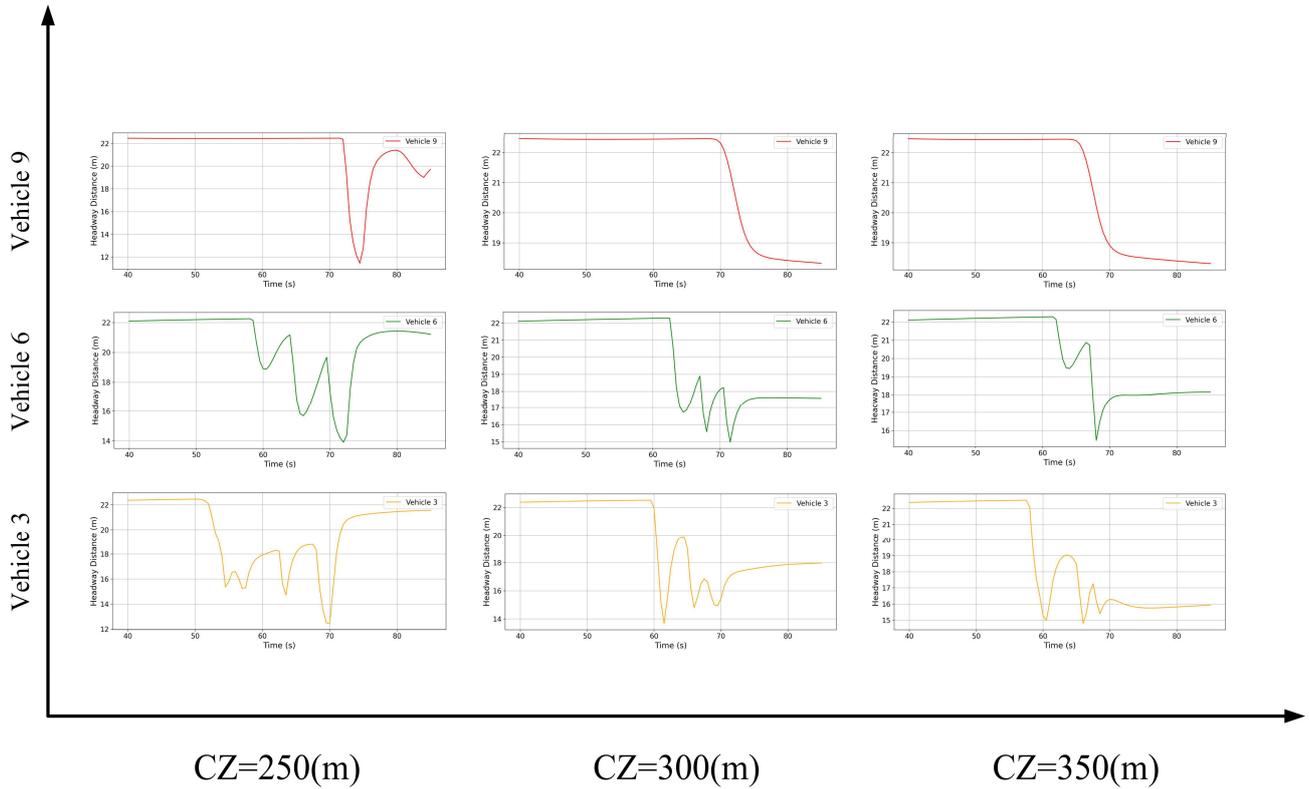

Fig.10. Graph showing the impact of different lengths of control zones (CZ) on the headway distance of vehicles in the platoon under the "1+n+1" mixed platoon control scenario. The yellow, green, and red curves represent the variation of the headway distance of the 3rd, 6th, and 9th vehicles in the platoon over time, respectively.

We also conducted targeted analysis on the length of the control zone (CZ), as shown in Figure 10, where curves depicting the variation of the headway distance of the 3rd, 6th, and 9th vehicles in the platoon, respectively, are provided for CZ = 250 m, CZ = 300 m, and CZ = 350 m. Given our earlier assumption that communication delay and packet loss are not considered throughout the control process, the length of the control zone does not affect the effectiveness of the central coordinated controller in transmitting information to vehicles. Here, we focus solely on whether vehicles in the platoon can stably travel near the stop line and the time required to achieve stability.

By observing the aforementioned figures, we can easily observe that when CZ = 250 m, the amplitude of the variation in the headway distance of the 3rd and 6th vehicles near the front and middle of the platoon respectively increases with time without convergence. This is because when the length of the control zone is short, the platoon often fails to form a stable state before the leading CAV passes through the control zone, while the trailing CAVs cannot directly influence the vehicles near the front. As a result, the entire platoon passing through the control zone fails to form a stable state, and the spacing between vehicles is relatively large, leading to relatively poor traffic efficiency.

However, when CZ = 300 m and CZ = 350 m, we observe that vehicles in different parts of the platoon can converge to a relatively stable headway distance at a faster rate. By comparison, when CZ = 350 m, vehicles in the platoon can begin adjusting their headway distance and achieve stability at an earlier stage, although the difference in the time it takes to form a stable platoon is negligible. One of the final control objectives of this study is for the platoon to achieve stability before reaching the stop line at the signalized intersection. Therefore, whether it happens a bit earlier or later has little impact on the control objective. However, in practice, excessively long control zones are impractical as drivers do not consider whether they can pass through the intersection when the traffic light is still far away.

In conclusion, too short control zones cannot achieve the control objectives of the "1+n+1" mixed traffic control method proposed in this paper, while excessively long control zones make it difficult to apply the theory to practice. Therefore, selecting an appropriate control zone length is essential.

# 6 Conclusions

The CAV control method proposed in this paper for mixed traffic flow through signalized intersections, considering the vehicle's backward-looking effect, namely the "1+n+1" mixed platoon control method, can effectively improve the maneuverability at signalized intersections. It offers greater advantages compared to traditional mixed traffic flow control methods that only consider the leading vehicle, such as higher traffic efficiency and lower fuel consumption. Through rigorous theoretical analysis and extensive simulation verification, this paper demonstrates that the proposed mixed platoon control method is controllable, stable, and has a certain universality. The optimal control framework of this paper is established based on considering the velocity deviation and fuel consumption of the entire mixed traffic flow, optimizing vehicle velocity when passing the stop line at signalized intersections to maximize traffic flow within a single green phase.

However, there are still some limitations in the current research. For example, it is difficult to shorten the length of the control zone as much as possible. Based on considering the vehicle's backward-looking effect, the time for the platoon to form stable travel is often long. Future improvements can be made by enhancing the vehicle-following models considering the backward-looking effect, allowing the platoon to reach a stable state more quickly, thus shortening the control zone. Additionally, the CAVs controlled in this paper are only the leading and trailing CAVs in the mixed traffic flow, while the vehicles guided in the middle of the mixed flow consist of both HDVs and CAVs. However, we have treated all guided vehicles as manually driven HDVs. In the future, employing cooperative control algorithms to simultaneously control multiple CAVs to further improve the overall performance of mixed traffic flow through intersections will be an interesting research topic. Lastly, this paper only focuses on the longitudinal control of vehicles and does not address lane-changing behavior. Resolving lane-changing issues for CAVs and HDVs is also an important research direction in the future.

**Acknowledgments**

This research was partly funded by National Natural Science Foundation of China (Grant Nos.71801149, 71801153, 52172371); Projects of Natural Science Foundation of Shanghai(20ZR1422300); Project of the Program for Shanghai Academic Research Leader (No. 21XD1401100) and Technical Service Platform for Vibration and Noise Testing and Control of New Energy Vehicles (Grant No. 18DZ2295900).